\documentclass[sigconf]{acmart}

\usepackage{url}
\usepackage{amsmath}
\usepackage{subfigure}
\usepackage{microtype}      
\usepackage{amsmath} 
\usepackage[ruled,vlined]{algorithm2e}
\usepackage{setspace}
\usepackage{graphicx}
\usepackage{wrapfig}
\usepackage{float}
\usepackage{caption}
\AtBeginDocument{%
  \providecommand\BibTeX{{%
    \normalfont B\kern-0.5em{\scshape i\kern-0.25em b}\kern-0.8em\TeX}}}

\copyrightyear{2022}
\acmYear{2022}
\setcopyright{acmcopyright}
\acmConference[ICPE '22 Companion]{Companion of the 2022 ACM/SPEC International Conference on Performance Engineering}{April 9--13, 2022}{Bejing, China}
\acmBooktitle{Companion of the 2022 ACM/SPEC International Conference on Performance Engineering (ICPE '22 Companion), April 9--13, 2022, Bejing, China}
\acmPrice{15.00}
\acmDOI{10.1145/3491204.3527462}
\acmISBN{978-1-4503-9159-7/22/04}

\acmSubmissionID{icpe102hot}

\begin{document}
\fancyhead{}

\title{MiSeRTrace: Kernel-level Request Tracing for Microservice Visibility}

\author{Thrivikraman V}
\authornotemark[1]
\affiliation{%
 \institution{PES University}
 \city{Bengaluru}
 \state{Karnataka}
 \country{India}}
\email{thrivikraman.vs@gmail.com}

\author{Vishnu R. Dixit}
\authornotemark[1]
\affiliation{%
 \institution{PES University}
 \city{Bengaluru}
 \state{Karnataka}
 \country{India}}
\email{rvdixit23@gmail.com}

\author{Nikhil Ram S}
\authornote{These authors contributed equally to this research.}
\affiliation{%
 \institution{PES University}
 \city{Bengaluru}
 \state{Karnataka}
 \country{India}}
\email{nikhilsram.off@gmail.com}

\author{Vikas K. Gowda}
\authornotemark[1]
\affiliation{%
 \institution{PES University}
 \city{Bengaluru}
 \state{Karnataka}
 \country{India}}
\email{vikasgowdaoffl@gmail.com}

\author{Santhosh Kumar Vasudevan}
\affiliation{%
 \institution{PES University}
 \city{Bengaluru}
 \state{Karnataka}
 \country{India}}
\email{santvasu@gmail.com}

\author{Subramaniam Kalambur}
\affiliation{%
 \institution{PES University}
 \city{Bengaluru}
 \state{Karnataka}
 \country{India}}
 \email{subramaniamkv@pes.edu}

\renewcommand{\shortauthors}{Thrivikraman, Vishnu, Nikhil Ram, Vikas, et al.}

\begin{abstract}
With the evolution of microservice applications, the underlying architectures have become increasingly complex compared to their monolith counterparts. This mainly brings in the challenge of observability. By providing a deeper understanding into the functioning of distributed applications, observability enables improving the performance of the system by obtaining a view of the bottlenecks in the implementation. The observability provided by currently existing tools that perform dynamic tracing on distributed applications is limited to the user-space and requires the application to be instrumented to track request flows. 
In this paper, we present a new open-source framework MiSeRTrace that can trace the end-to-end path of requests entering a microservice application at the kernel space without requiring instrumentation or modification of the application. Observability at the comprehensiveness of the kernel space allows breaking down of various steps in activities such as network transfers and IO tasks, thus enabling root cause based performance analysis and accurate identification of hotspots. MiSeRTrace supports tracing user-enabled kernel events provided by frameworks such as bpftrace or ftrace and isolates kernel activity associated with each application request with minimal overheads. We then demonstrate the working of the solution with results on a benchmark microservice application.
\end{abstract}



\maketitle

\section{Introduction}
Most of the recent client-server web applications are adopting microservice architectures due to their benefits like modularity and scalability \cite{9251239} over their monolith counterparts \cite{7333476}. The highly networked microservice architecture brings in challenges of visibility into the underlying functioning of the application, introduces overheads \cite{7581269}, and influences server design \cite{8362750}. Our solution enables the required observability of the application through dynamic tracing at the kernel space. Dynamic tracing is the process of tracking the end-to-end path of user requests from the time the request hits the load balancer of the application until a response to that request is sent out. MiSeRTrace\footnote[1]{https://github.com/MiSeRTrace/MiSeRTrace} ({\bf Mi}cro{\bf Se}rvice {\bf R}equest {\bf Trace}) provides insights into how every client request is serviced in the kernel which aids in identifying and understanding performance differentials.
\par Containers in a microservice application communicate with other containers through lightweight API calls which are internally TCP network transfers. Every container in a microservice application is a process, and this process forks many threads to handle the incoming requests. MiSeRTrace primarily tracks these events in order to isolate the path of every client request through the application. This level of monitoring avoids the requirement of tracing libraries such as the OpenTracing API \cite{OpenTracing} for instrumentation or proxy sidecars like Istio \cite{istio}. The tool utilizes certain static and dynamic kernel tracepoints to track the above events. Monitoring these tracepoints is enabled by kernel tracing utilities such as bpftrace \cite{bpftrace} and ftrace \cite{ftrace}. Bpftrace is a high level tracing language based on extended Berkeley Packet Filter (eBPF) \cite{eBPF}, which allows programs to execute sandboxed code in the kernel. Ftrace is a tracing framework that is built into the Linux kernel and provides visibility primarily into kernel functions and static events. Henceforth, we will refer to the data captured by these frameworks as trace logs.
\par Our implementation supports both bpftrace and ftrace as tracing backends. Apart from tracepoints utilized by our framework, users can enable any of the tracepoints and features provided by bpftrace/ftrace that they wish to monitor. MiSeRTrace initially identifies the request spans, i.e. the duration spent by a particular thread on servicing a client request or a subsequent internal request. It then associates all these request spans with unique traces. MiSeRTrace subsequently buckets all the time-stamped user-enabled trace logs into the request spans that triggered them to assist in latency estimate studies.

\section{Related Work}
When it comes to dynamic tracing, there are various tools that are widely used to monitor hotspots in the application. The OpenTracing API is one such framework that utilizes Trace IDs and Span IDs to achieve application instrumentation. A trace is used to uniquely identify all operations performed for an incoming client request, and a span is used to refer to a single operation within a trace. Jaeger \cite{Jaeger} and Zipkin \cite{Zipkin} are a couple of dynamic request tracers that utilize the Opentracing API. They are open source tracing tools that are used for monitoring and troubleshooting large scale distributed microservice applications. Kieker \cite{HASSELBRING2020100019} is another tracing software used to monitor and analyze the performance of monolithic and microservice applications and performs request tracing with the help of measurement probes. Dapper \cite{36356}, which is Google’s distributed system tracing infrastructure, relies on the instrumentation of RPC libraries, control flow libraries, and threading. Dapper is capable of identifying distributed control paths with almost no intervention from the developers.
\par These tools make use of low latency datastores where the data is frequently pushed to and the developers can gather this information generally through a web based interface. However, they provide insights that are limited to the user space i.e they provide information about the path of the request through the various containers in the microservice and corresponding latency estimates. All kernel activity as a consequence of the client request is mostly unobserved. These frameworks also require the application being monitored or underlying libraries to be instrumented. Both the above issues are overcome by MiSeRTrace by providing the capacity to observe all kernel activity of an un-instrumented application.

\section{Architecture}
\begin{figure}[h]
    \centering
    \includegraphics[width=8cm]{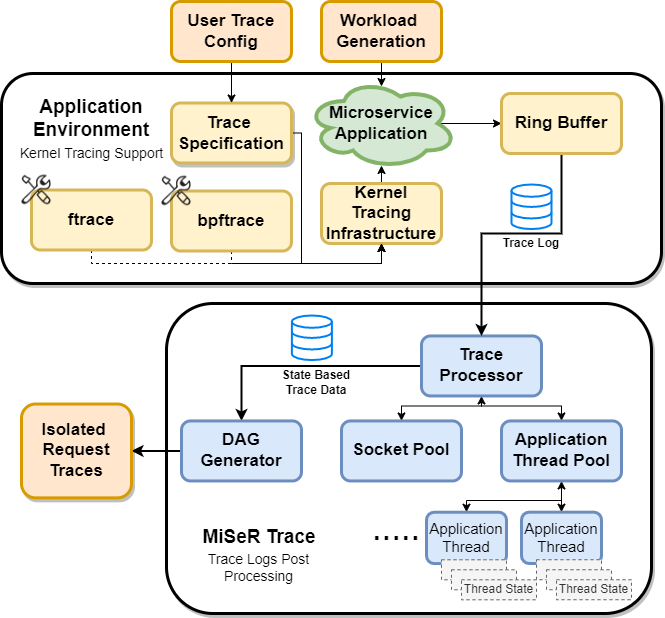}
    \caption{MiSeRTrace workflow}
\end{figure}
The end-to-end usage and working of MiSeRTrace involves 4 main steps which are broadly illustrated in Figure 1.
\par {\bf (1) Trace Specification:}
MiSeRTrace currently supports bpftrace and ftrace as tracing backends. Events built into the kernel, tracers built into ftrace, custom user probes and events are all supported. The events required by the tool to trace requests through the application are enabled by default. The user can specify a trace configuration to observe kernel activity beyond the default setting. Trace environment specifications like PIDs of the application to be tracked, tracking of forked processes, clocks, and buffer sizes can be configured.
\par {\bf (2) Workload Generation and Tracing of the containerized microservice application:}
The specified kernel activities are efficiently recorded into the bpftrace/ftrace in-memory ring buffer during the workload execution. On tracing the select events required by MiSeRTrace, there is a reduction of request throughput by around 5-6\%. Further, kernel events can be enabled based on the user’s interest. MiSeRTrace post-processes the trace logs and hence does not add to the overhead. Upon completion of the workload, the trace logs generated are passed onto the tool.
\par {\bf (3) Processing of the Trace Logs:}
MiSeRTrace performs sequential post processing on the kernel trace logs. Upon processing, the generated data consists of all the request spans, where spans are represented as states of the application threads as we will explain in the Thread State Model (TSM). All kernel events that were enabled by the trace specification are associated with the respective states.
\par {\bf (4) Generation of Isolated Request Traces:}
Once the trace logs have been completely processed, the Directed Acyclic Graph (DAG) Generator consolidates all the states associated with each unique client request and isolates them into time ordered traces. Each request trace is a DAG that is generated by a depth-first recursive algorithm.

MiSeRTrace is designed and implemented as the following 3 hierarchical cohesive entities.
\par {\bf (1) Trace Processor:}
The trace processor is responsible for the creation of thread pools and socket pools maintained in MiSeRTrace. The kernel trace logs are passed into the trace processor ordered by time of occurrence. Generation of trace IDs that uniquely identify the client requests is also handled here. The trace logs are validated, preprocessed, and sent to the thread processor. Once all the trace logs are processed by the lower levels, the trace processor transforms the state-based trace data into a representation suited for the DAG Generator.
\par {\bf (2) Thread Processor:}
The thread processor maintains the thread pool created by the trace processor. It tracks and creates representations of threads of the application brought about by forking of threads during workload runs. The threads are maintained in two pools, a pool of active threads and a pool of terminated threads.
\par {\bf (3) Thread State Processor and Thread State Model (TSM):}
Each thread object has its own State Processor which processes all trace records associated with that thread, updates states as per the TSM, and also propagates the trace ID through the request lifecycle. TSM is motivated by a need to associate steps within the request lifecycle to application threads. As per the model, there are two types of states depicted in Figure 2, namely Network Thread States and Fork Thread States. Each state is associated with only a single application thread.\\
\begin{wrapfigure}{l}{0.25\textwidth}
  \begin{center}
    \includegraphics[width=0.25\textwidth]{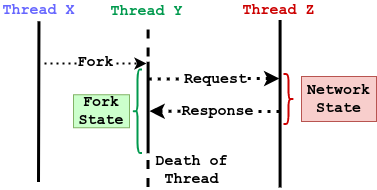}
  \end{center}
  \caption{Types of Thread States}
\end{wrapfigure}
\textbf{\textit{Network Thread States:}} They are created to track incoming TCP requests to a thread. A thread has one state for every incoming request. Each network state in a thread can be uniquely identified by a source thread, TCP 4 tuple, and the trace ID of the incoming request. The source thread refers to the thread which is sending data. Any communication between two threads involves the usage of a socket which is identified by the TCP 4 tuple which consists of the IP addresses and ports of both the sender and receiver. Each socket is transmitting either an application REQUEST or RESPONSE. The span of this state is from the time of request arrival to the time a response is sent for that request.
\\\textbf{\textit{Fork Thread States:}} They are created when a thread forks to handle multiple incoming requests or send multiple asynchronous requests to other threads. Each state is uniquely identified by the parent thread and the trace ID of the request which triggered the parent thread to spawn the child thread. The span of this state is from the time the child thread is created to the time it terminates.

\section{Implementation}
\begin{algorithm}[h]
\SetAlgoLined
ThreadPool $\leftarrow$ Map[Pid$\rightarrow$Thread]\\
SocketPool $\leftarrow$ Map[(IpPair, PortPair)$\rightarrow$socket]\\
SendSyscallsSet $\leftarrow$ Set\{sendto, sendmsg, write, writev\}\\
RecieveSyscallsSet $\leftarrow$ Set\{recvfrom, recvmsg, read, readv\}\\
NetworkStateStore $\leftarrow$ Map[(SrcThread, tcp4tuple, TraceId) $\rightarrow$ NetworkState] // one store per thread \\
\uIf{currentSyscall $\in$ SendSyscallsSet} {
    \If{currentTracepoint.event = tcpSendKprobe} {
        senderSock $\leftarrow$ SocketPool.get(currentTracepoint.tcp4tuple)\\
        senderSock.senderThread $\leftarrow$  currentThread\\
        senderSock.type $\leftarrow$  REQUEST\\
        \textit{/*To check if a Response is being sent*/}\\
        \ForAll{state $\in$  NetworkStateStore.values()} {
            \textit{/*Source and destination here refer to the pairs (sourceIp, sourcePort) and (destinationIp, destinationPort) respectively*/}\\
            \If{state.source =  senderSock.destination} {
                senderSock.type $\leftarrow$  RESPONSE\\
                state.endSpan()\\
                break
            }
        }
    }
}
\ElseIf{ currentSyscall $\in$  RecieveSyscallsSet } {
    \If{ currentTracepoint.event =  tcpRcvSpaceAdjust } {
         recieverSock $\leftarrow$  SocketPool.get(currentTracepoint.tcp4tuple) \\
        \If{ receiverSock.type =  REQUEST } {
		     senderThread $\leftarrow$  receiverSock.senderThread \\
		     \textit{/*All active threadStates of the senderThread are propogated to the currentThread*/}\\
	        \ForAll{ traceID $\in$  senderThread.allTraces } {
		         stateKey $\leftarrow$  (senderThread, receiverSock.tcp4tuple, traceId)\\
	             state $\leftarrow$  createNetworkState(stateKey)\\
	             state.source  $\leftarrow$ receiverSock.source\\
			     state.startSpan()\\
			     NetworkStateStore[stateKey] $\leftarrow$  state 
            }
        }
    }
}
\caption{TSM algorithm to handle network transfers}
\end{algorithm}
MiSeRTrace has been implemented to allow flexibility in terms of intrusiveness by giving the user control over the amount of tracing.  In order to minimize the overhead caused by tracing, only the PIDs that belong to the application are traced. This is done by examination of the PIDs present on the containerization engine’s network and passing on the same to the tracing backends (bpftrace/ftrace). The size of the in-memory ring buffer is to be set depending on how much kernel activity is being captured to prevent loss of data in the trace logs.
\par The basis of tracing request flows through the application is by monitoring a set of system calls activated when sending and receiving messages (SendSyscallsSet and RecieveSyscallsSet in Algorithm 1), and kernel events that occur within these system calls. These are available as static tracepoints in the tracing backends. In addition to these, a custom kprobe required by MiSeRTrace is inserted into the kernel. This probe is associated with the sock\_sendmsg and \_\_sys\_sendmsg functions and is used to detect data being sent over a TCP connection after necessary permissions are acquired. Custom probes such as these can also be added by the user to gain a view of kernel activity at maximum configurability. The tracing backends have respective provisions to add and activate such probes. To capture the reception of data over a TCP connection, a static tracepoint called tcp\_rcv\_space\_adjust is used, which is triggered every time data is copied to the user space at the receiver’s end. 
A total of 21 events are utilized by MiSeRTrace. The information provided by these events is used to update the thread states of the TSM in order to trace the request flow.
Algorithm 1 is a simplified version of the process of creating and updating the Network Thread States for any given thread by monitoring the networking events. A similar algorithm also manages the Fork Thread States by utilizing the scheduling events - sched\_process\_fork and sched\_process\_exit.
\par While using ftrace, the logistics of managing the trace data is handled by trace-cmd \cite{tracecmd}, an interface for ftrace. In the case where bpftrace is used as the backend, there are a few extra steps to ensure while collecting trace logs. Bpftrace does not notify when space for the ring buffer is allocated or when the trace logs have been written to disk and hence these need to be managed by the user. The logs obtained from bpftrace are per-CPU time ordered logs but are not time-ordered globally across CPUs. MiSeRTrace also encompasses a sorting system for the ingestion of such trace logs from bpftrace.

\section{Results}
MiSeRTrace was put to use on an open-source microservice benchmark suite known as DeathStarBench \cite{10.1145/3297858.3304013}, which consists of many end-to-end services with representative workloads. The social network application is one of the benchmarks in DeathStarBench, implemented with loosely-coupled containerized \cite{Docker} microservices communicating with each other via Thrift RPCs. It primarily consists of 3 levels - the front end (load balancer, Nginx), the logic, and the backend stores (Memcached, MongoDB, Redis).
\par The benchmark was run on a system with two CPU sockets, where each socket is an AMD EPYC 7401 24-Core Processor. Each core has 2 logical CPUs, and hence the machine has a total of 96 CPUs. Each socket consists of 4 NUMA nodes. This non-uniform memory access configuration where local and remote memory accesses take different times was used to simulate a distributed cluster-like environment. This server runs Linux kernel 5.4.0 and has a working memory of 128 GB.
\par A workload was generated on the benchmark using wrk \cite{wrk}, a HTTP benchmarking utility. Any form of statistical/state-based analysis using kernel events can be performed by feeding stubs of code into MiSeRTrace. As an example, the trace-points sched: sched\_migrate\_task and exceptions:page\_fault\_user were enabled for monitoring with the objective of counting the number of occurrences of these events. These trace logs were then processed by MiSeRTrace which produced the request flow DAGs for all client requests, one of which is represented in Figure 3. Each segment here represents a state as defined by the Thread State Model, whose span is represented by the length of the segment. Each state in the figure shows the PID of the associated thread and the name of the microservice running on it. In Figure 3, PID 2066822 forks PID 2066823 which sends a TCP request to PID 1966384. The span of a state is concluded upon sending a response/death of the thread. One of the insights that can be derived from this is that the number of page faults that occurred in the Home-Timeline Redis span is comparatively higher. These statistics can also be compared across multiple request traces to understand the performance differentials.

\section{Conclusion and Future Work}
\begin{figure}[h]
    \centering
    \includegraphics[width=\linewidth]{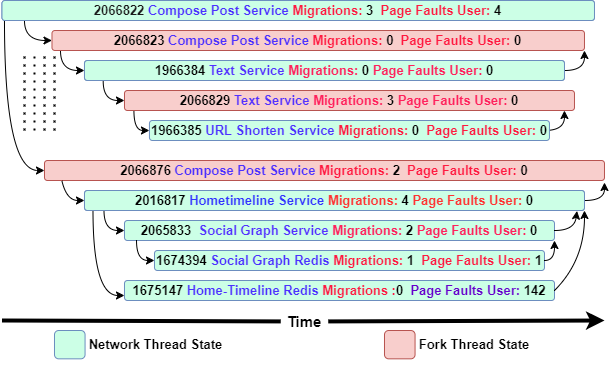}
    \captionsetup{justification=centering}
    \caption{A representative section of states of a single client request trace along with the tally of the monitored events}
\end{figure}
We have presented MiSeRTrace, a framework for tracing requests to microservice applications at the kernel space. It is capable of monitoring an un-instrumented application with minimal overheads. The tool includes provisions for enabling all the features and events of tracing backends such as bpftrace/ftrace. Subsequently, we demonstrated the usage of MiSeRTrace to trace the end-to-end path of client requests and user enabled events on a benchmark microservice application. With the exhaustive observability brought about by this approach, kernel-based application optimizations become much more accessible. In this paper, our experiments were conducted on a large core-count server. As future work, MiSeRTrace can also be extended to support tracing of kernel activity on a cluster of machines to derive kernel insights irrespective of the scale of the application.

\bibliographystyle{ACM-Reference-Format}
\bibliography{sample-base}


\begin{thebibliography}{17}


\ifx \showCODEN    \undefined \def \showCODEN     #1{\unskip}     \fi
\ifx \showDOI      \undefined \def \showDOI       #1{#1}\fi
\ifx \showISBNx    \undefined \def \showISBNx     #1{\unskip}     \fi
\ifx \showISBNxiii \undefined \def \showISBNxiii  #1{\unskip}     \fi
\ifx \showISSN     \undefined \def \showISSN      #1{\unskip}     \fi
\ifx \showLCCN     \undefined \def \showLCCN      #1{\unskip}     \fi
\ifx \shownote     \undefined \def \shownote      #1{#1}          \fi
\ifx \showarticletitle \undefined \def \showarticletitle #1{#1}   \fi
\ifx \showURL      \undefined \def \showURL       {\relax}        \fi
\providecommand\bibfield[2]{#2}
\providecommand\bibinfo[2]{#2}
\providecommand\natexlab[1]{#1}
\providecommand\showeprint[2][]{arXiv:#2}

\bibitem[bpf(n d)]%
        {bpftrace}
 \bibinfo{year}{[n. d.]}\natexlab{}.
\newblock \bibinfo{booktitle}{\emph{bpftrace}}.
\newblock
\urldef\tempurl%
\url{https://github.com/iovisor/bpftrace}
\showURL{%
\tempurl}


\bibitem[Doc(n d)]%
        {Docker}
 \bibinfo{year}{[n. d.]}\natexlab{}.
\newblock \bibinfo{booktitle}{\emph{Docker}}.
\newblock
\urldef\tempurl%
\url{https://www.docker.com}
\showURL{%
\tempurl}


\bibitem[eBP(n d)]%
        {eBPF}
 \bibinfo{year}{[n. d.]}\natexlab{}.
\newblock \bibinfo{booktitle}{\emph{eBPF}}.
\newblock
\urldef\tempurl%
\url{https://ebpf.io}
\showURL{%
\tempurl}


\bibitem[ftr(n d)]%
        {ftrace}
 \bibinfo{year}{[n. d.]}\natexlab{}.
\newblock \bibinfo{booktitle}{\emph{ftrace}}.
\newblock
\urldef\tempurl%
\url{https://www.kernel.org/doc/Documentation/trace/ftrace.txt}
\showURL{%
\tempurl}


\bibitem[ist(n d)]%
        {istio}
 \bibinfo{year}{[n. d.]}\natexlab{}.
\newblock \bibinfo{booktitle}{\emph{Istio}}.
\newblock
\urldef\tempurl%
\url{https://istio.io/latest/about/service-mesh/}
\showURL{%
\tempurl}


\bibitem[Jae(n d)]%
        {Jaeger}
 \bibinfo{year}{[n. d.]}\natexlab{}.
\newblock \bibinfo{booktitle}{\emph{Jaeger}}.
\newblock
\urldef\tempurl%
\url{https://www.jaegertracing.io}
\showURL{%
\tempurl}


\bibitem[Ope(n d)]%
        {OpenTracing}
 \bibinfo{year}{[n. d.]}\natexlab{}.
\newblock \bibinfo{booktitle}{\emph{OpenTracing}}.
\newblock
\urldef\tempurl%
\url{https://opentracing.io}
\showURL{%
\tempurl}


\bibitem[tra(n d)]%
        {tracecmd}
 \bibinfo{year}{[n. d.]}\natexlab{}.
\newblock \bibinfo{booktitle}{\emph{trace-cmd}}.
\newblock
\urldef\tempurl%
\url{https://trace-cmd.org/}
\showURL{%
\tempurl}


\bibitem[wrk(n d)]%
        {wrk}
 \bibinfo{year}{[n. d.]}\natexlab{}.
\newblock \bibinfo{booktitle}{\emph{wrk}}.
\newblock
\urldef\tempurl%
\url{https://github.com/wg/wrk}
\showURL{%
\tempurl}


\bibitem[Zip(n d)]%
        {Zipkin}
 \bibinfo{year}{[n. d.]}\natexlab{}.
\newblock \bibinfo{booktitle}{\emph{Zipkin}}.
\newblock
\urldef\tempurl%
\url{https://zipkin.io}
\showURL{%
\tempurl}


\bibitem[Caculo et~al\mbox{.}(2020)]%
        {9251239}
\bibfield{author}{\bibinfo{person}{Sriyash Caculo}, \bibinfo{person}{Kanishka
  Lahiri}, {and} \bibinfo{person}{Subramaniam Kalambur}.}
  \bibinfo{year}{2020}\natexlab{}.
\newblock \showarticletitle{Characterizing the Scale-Up Performance of
  Microservices using TeaStore}. In \bibinfo{booktitle}{\emph{2020 IEEE
  International Symposium on Workload Characterization (IISWC)}}.
  \bibinfo{pages}{48--59}.
\newblock
\urldef\tempurl%
\url{https://doi.org/10.1109/IISWC50251.2020.00014}
\showDOI{\tempurl}


\bibitem[Gan and Delimitrou(2018)]%
        {8362750}
\bibfield{author}{\bibinfo{person}{Yu Gan} {and} \bibinfo{person}{Christina
  Delimitrou}.} \bibinfo{year}{2018}\natexlab{}.
\newblock \showarticletitle{The Architectural Implications of Cloud
  Microservices}.
\newblock \bibinfo{journal}{\emph{IEEE Computer Architecture Letters}}
  \bibinfo{volume}{17}, \bibinfo{number}{2} (\bibinfo{year}{2018}),
  \bibinfo{pages}{155--158}.
\newblock
\urldef\tempurl%
\url{https://doi.org/10.1109/LCA.2018.2839189}
\showDOI{\tempurl}


\bibitem[Gan et~al\mbox{.}(2019)]%
        {10.1145/3297858.3304013}
\bibfield{author}{\bibinfo{person}{Yu Gan}, \bibinfo{person}{Yanqi Zhang},
  \bibinfo{person}{Dailun Cheng}, \bibinfo{person}{Ankitha Shetty},
  \bibinfo{person}{Priyal Rathi}, \bibinfo{person}{Nayan Katarki},
  \bibinfo{person}{Ariana Bruno}, \bibinfo{person}{Justin Hu},
  \bibinfo{person}{Brian Ritchken}, \bibinfo{person}{Brendon Jackson},
  \bibinfo{person}{Kelvin Hu}, \bibinfo{person}{Meghna Pancholi},
  \bibinfo{person}{Yuan He}, \bibinfo{person}{Brett Clancy},
  \bibinfo{person}{Chris Colen}, \bibinfo{person}{Fukang Wen},
  \bibinfo{person}{Catherine Leung}, \bibinfo{person}{Siyuan Wang},
  \bibinfo{person}{Leon Zaruvinsky}, \bibinfo{person}{Mateo Espinosa},
  \bibinfo{person}{Rick Lin}, \bibinfo{person}{Zhongling Liu},
  \bibinfo{person}{Jake Padilla}, {and} \bibinfo{person}{Christina
  Delimitrou}.} \bibinfo{year}{2019}\natexlab{}.
\newblock \showarticletitle{An Open-Source Benchmark Suite for Microservices
  and Their Hardware-Software Implications for Cloud \& Edge Systems}. In
  \bibinfo{booktitle}{\emph{Proceedings of the Twenty-Fourth International
  Conference on Architectural Support for Programming Languages and Operating
  Systems}} (Providence, RI, USA) \emph{(\bibinfo{series}{ASPLOS '19})}.
  \bibinfo{publisher}{Association for Computing Machinery},
  \bibinfo{address}{New York, NY, USA}, \bibinfo{pages}{3–18}.
\newblock
\showISBNx{9781450362405}
\urldef\tempurl%
\url{https://doi.org/10.1145/3297858.3304013}
\showDOI{\tempurl}


\bibitem[Hasselbring and {van Hoorn}(2020)]%
        {HASSELBRING2020100019}
\bibfield{author}{\bibinfo{person}{Wilhelm Hasselbring} {and}
  \bibinfo{person}{André {van Hoorn}}.} \bibinfo{year}{2020}\natexlab{}.
\newblock \showarticletitle{Kieker: A monitoring framework for software
  engineering research}.
\newblock \bibinfo{journal}{\emph{Software Impacts}}  \bibinfo{volume}{5}
  (\bibinfo{year}{2020}), \bibinfo{pages}{100019}.
\newblock
\showISSN{2665-9638}
\urldef\tempurl%
\url{https://doi.org/10.1016/j.simpa.2020.100019}
\showDOI{\tempurl}


\bibitem[Sigelman et~al\mbox{.}(2010)]%
        {36356}
\bibfield{author}{\bibinfo{person}{Benjamin~H. Sigelman},
  \bibinfo{person}{Luiz~André Barroso}, \bibinfo{person}{Mike Burrows},
  \bibinfo{person}{Pat Stephenson}, \bibinfo{person}{Manoj Plakal},
  \bibinfo{person}{Donald Beaver}, \bibinfo{person}{Saul Jaspan}, {and}
  \bibinfo{person}{Chandan Shanbhag}.} \bibinfo{year}{2010}\natexlab{}.
\newblock \bibinfo{booktitle}{\emph{Dapper, a Large-Scale Distributed Systems
  Tracing Infrastructure}}.
\newblock \bibinfo{type}{{T}echnical {R}eport}. \bibinfo{institution}{Google,
  Inc.}
\newblock
\urldef\tempurl%
\url{https://research.google.com/archive/papers/dapper-2010-1.pdf}
\showURL{%
\tempurl}


\bibitem[Ueda et~al\mbox{.}(2016)]%
        {7581269}
\bibfield{author}{\bibinfo{person}{Takanori Ueda}, \bibinfo{person}{Takuya
  Nakaike}, {and} \bibinfo{person}{Moriyoshi Ohara}.}
  \bibinfo{year}{2016}\natexlab{}.
\newblock \showarticletitle{Workload characterization for microservices}. In
  \bibinfo{booktitle}{\emph{2016 IEEE International Symposium on Workload
  Characterization (IISWC)}}. \bibinfo{pages}{1--10}.
\newblock
\urldef\tempurl%
\url{https://doi.org/10.1109/IISWC.2016.7581269}
\showDOI{\tempurl}


\bibitem[Villamizar et~al\mbox{.}(2015)]%
        {7333476}
\bibfield{author}{\bibinfo{person}{Mario Villamizar}, \bibinfo{person}{Oscar
  Garcés}, \bibinfo{person}{Harold Castro}, \bibinfo{person}{Mauricio Verano},
  \bibinfo{person}{Lorena Salamanca}, \bibinfo{person}{Rubby Casallas}, {and}
  \bibinfo{person}{Santiago Gil}.} \bibinfo{year}{2015}\natexlab{}.
\newblock \showarticletitle{Evaluating the monolithic and the microservice
  architecture pattern to deploy web applications in the cloud}. In
  \bibinfo{booktitle}{\emph{2015 10th Computing Colombian Conference (10CCC)}}.
  \bibinfo{pages}{583--590}.
\newblock
\urldef\tempurl%
\url{https://doi.org/10.1109/ColumbianCC.2015.7333476}
\showDOI{\tempurl}


\end{thebibliography}

\end{document}